# Title

ROSpace: Intrusion Detection Dataset for a ROS2-Based Cyber-Physical System

# Authors

Tommaso Puccetti[1], Simone Nardi[2], Cosimo Cinquilli[1], Tommaso Zoppi[3], Andrea Ceccarelli[1]

**Affiliations**

1. Department of Mathematics and Informatics, University of Florence, Viale Morgagni 67/a, 50134 Firenze (FI), Italy.
2. Mermec Engineering, Via Livornese 1019, 56122, Pisa, Italy.
3. Department of Engineering and Information Science, University of Trento, Via Sommarive 9, 38124, Trento, Italy.

corresponding author(s): Tommaso Puccetti (tommaso.puccetti@unifi.it).

# Abstract

Most of the intrusion detection datasets to research machine learning-based intrusion detection systems (IDSs) are devoted to cyber-only systems, and they typically collect data from one architectural layer. Additionally, often the attacks are generated in dedicated attack sessions, without reproducing the realistic alternation and overlap of normal and attack actions. We present a dataset for intrusion detection by performing penetration testing on an embedded cyber-physical system built over Robot Operating System 2 (ROS2). Features are monitored from three architectural layers: the Linux operating system, the network, and the ROS2 services. The dataset is structured as a time series and describes the expected behavior of the system and its response to ROS2-specific attacks: it repeatedly alternates periods of attack-free operation with periods when a specific attack is being performed. Noteworthy, this allows measuring the time to detect an attacker and the number of malicious activities performed before detection. Also, it allows training an intrusion detector to minimize both, by taking advantage of the numerous alternating periods of normal and attack operations.

# Background & Summary

*Intrusion Detection Systems* (IDS[9]) are of critical importance to secure systems against attacks. They consist of two major components: a monitor and a detector. The first collects data from a running system, while the second analyses these data to discern between legitimate (normal) or malicious behaviors. While most commercial intrusion detectors rely on static rules, anomaly detection[42] approaches based on machine learning constitute an attractive and widely researched alternative. They can develop complex decision criteria to classify each data point either as *normal* or *attack*, with also the ability to identify zero-days[18], i.e., attacks never observed before and that are not yet described by static rules. An anomaly-based intrusion detector finds patterns in data that do not conform to the expected behavior of a system (or a network): these patterns, called anomalies[42], are suspected attacks. Clearly, anomaly-based intrusion detectors are built under the assumption that ongoing attacks generate observable anomalies in the trend of the monitored performance indicators, or features, of the system[32], network[45], or both[44].

To successfully train and test an anomaly-based intrusion detector, it is important to use realistic datasets specific to the operational domain and with analogous features and distribution concerning real-world data. The intrusion detection datasets available at the state-of-the-art mostly focus on cyber-only systems, while examples with cyber-physical systems are exceedingly scarce. In addition, these datasets are usually focused on monitoring the network traffic, without providing useful data to observe the manifestation of anomalies in other architectural layers. Also, the network traffic in the absence of attacks is often generated by modeling user profiles and simulating the user's behavior: the normal and the attack behavior are unrelated because they are collected separately, ultimately preventing the datasets from capturing the dynamics of transiting from an attack-free state to an under-attack state. These pitfalls evidence the lack of intrusion detection datasets that accurately describe the behavior of a cyber-physical system in a real execution environment.

In this paper, we introduce the ROSPaCe (Robot Operating System 2 for Smart Passenger Center) dataset for intrusion detection, which is specific to the SPaCe (Smart Passenger Center) embedded cyber-physical system and aims to overcome the limitations stated above.

The SPaCe system orchestrates onboard and ground components of a tramline carriage to optimize public mobility. It collects data from onboard cameras and processes them to improve the quality of the service, for example facilitating fleet optimization, supporting passengers, and commanding maintenance interventions. The system is built using ROS2, a popular protocol for robotic and distributed systems [19,20]. The operational capabilities of the SPaCe system impose the application of security solutions. While IDSs are particularly suitable for the context, they cannot be viably applied without relevant training data.

For these reasons, we present the ROSPaCe dataset, which is specific to SPaCe and ROS2. We perform attacks through the execution of discovery and DoS attacks, for a total of 6 attacks, with 3 of them specific to ROS2. We collect data from the network interfaces, the operative system, and ROS2, and we merge the observations in a unique dataset using the timestamp. We label each data point indicating if it is recorded during the normal (attack-free) operation, or while the system is under attack. The dataset is organized as a time series in which we alternate sequences of normal (attack-free) operations, and sequences when attacks are carried out in addition to the normal operations. The goal of this strategy is to reproduce multiple scenarios of an attacker trying to penetrate the system.

The final version of ROSPaCe includes above 30 million data points with 482 features, collected over 4 days of operation and for a total of approximately 40.5 GBs. We also present a simplified version of the dataset, which differs in the number of features and consequently has a reduced dimension of 13.5 GBs. We evaluate the performances of the IDSs using the accuracy, ROC curve, and precision-recall curve. Also, we show how ROSPaCe can be exploited to measure the attack latency at the cost of a maximum acceptable false alarm rate. In fact, in practice, a system developer is interested in minimizing the attack latency (identify an attacker as soon as possible), while maintaining false alarms reasonably low to avoid too many unnecessary attack responses. This is not achievable with most of the state-of-the-art datasets, because the dynamics of transiting from an attack-free state to an under-attack state are insufficiently captured.

In summary, the relevant features of ROSPaCe are: i) it focuses on a cyber-physical system, being the first intrusion detection dataset targeting ROS2; ii) it monitors multiple architectural layers, not being limited to logging network traffic; iii) it is recorded from a real system under execution, so that there is no simulated traffic or simulated user's profile; iv) it is composed of multiple distinguishable scenarios, modeling an attacker attempting to penetrate a clean system while it is operating. The validation results demonstrate that the dataset is a valuable resource for training and evaluating IDSs for embedded cyber-physical systems and can be used as a landmark to develop and evaluate IDS solutions specific to ROS2-based systems.

Noteworthy, the attacks successfully targeted the system and degraded the system's performance: we remark that the SPaCe system has been successively modified to implement defenses, and the attacks are no longer effective against the deployed system.

## Related Works: Intrusion Detection Datasets

In this section, we report on the most used public datasets for intrusion detection with the help of Table 1, and we elaborate on the differences from ROSPaCe.

*KDD Cup99*[12] is a labeled dataset that collects the monitoring activity of a LAN architecture of a small US Air Force base. The dataset models the normal activity of the LAN and its behavior in several simulated intrusion scenarios. The total size of the datasets is 4 GBs for a total of almost 5 million records and 41 features describing the network communication. Even though the dataset is over 20 years old, it remains in use as a benchmark within the IDS research community[11].

*NSL-KDD*[12] is based on the KDD dataset and aims to solve some limitations of the original dataset. In particular, the authors eliminate duplicate records that prevent machine learning methods from discerning between irregular and benign instances. The NSL-KDD dataset comprises 22 attacks and 41 features, as in KDD Cup99.

Researchers at the Australian Defence Force Academy created two datasets, namely *ADFA-LD*[13] and *ADFA-WD*[13]. The datasets include system call traces collected from, respectively, Linux and Windows host machines. ADFA-LD is collected by executing different Internet services on the Ubuntu 11.04 operative system. It provides traces relative to normal behavior and 6 web-based attacks. ADFA-WD applies the same methodology to collect traces from web service applications on a Windows XP SP2 host, collecting traces of 12 web-based attacks that target known vulnerabilities of the host. The features are extracted semantically from the trace of executed system calls using natural language methods[37,38].

*ADFA-Netflow-IDS*[41] (simply termed ADFANet in the rest of the paper) is obtained by monitoring the network traffic generated by Cisco routers relying on the built-in NetFlows monitor. The fields provided by NetFlow are directly used as features for the dataset.

*ISCX2012*[15] collects observations from a real live network environment that relies on two synthetic profiles to generate traffic: one to simulate the attacks, and the other to simulate the normal traffic which includes HTTP, SMTP, POP3, SSH, and IMAP protocols. The dataset includes DoS and brute force attacks, providing 20 features extracted from the network.

*CICIDS17*[16] is an extension of the ISCX 2012 dataset that increases the number of features to 80 and introduces the profile technique to generate normal traffic. The dataset includes a broad range of attacks such as Brute Force FTP, Brute Force SSH, DoS, Heartbleed, Web Attacks, Infiltration, Botnet, and DDoS. The data are collected by monitoring a real network topology which includes modems, firewalls, switches, routers, and nodes with different operating systems: Microsoft Windows like Windows 10, Windows 8, Windows 7, and Windows XP, Apple macOS iOS, and Linux.

*CSE-CIC-IDS2018*[16] (simply termed CICIDS18) is realized by the Communication

| dataset name | year | environment | normal traffic | features type | number of features | millions of data points | application-specific attacks | attacker dynamics | accuracy |
|---|---|---|---|---|---|---|---|---|---|
| KDDCup99 | 1999 | real | conventional | network | 41 | 4.9 | no | yes | 0.85 |
| NSL-KDD | 2009 | real | conventional | network | 41 | 4.9 | no | yes | 0.84 |
| ADFA-LD | 2014 | real | conventional | syscall | 3 792 | 2.7 | yes | no | 0.90 |
| ADFA-WD | 2014 | real | conventional | syscall | 4 801 | 205.6 | yes | no | 0.91 |
| ADFA-Net | 2012 | real | conventional | network | 5 | 2.6 | no | yes | 0.89 |
| ISCX2012 | 2012 | real | simulated | network | 20 | 0.6 | no | yes | 0.87 |
| CICIDS17 | 2017 | real | simulated | network | 80 | 0.2 | no | yes | 0.89 |
| CICIDS18 | 2018 | virtualized | simulated | network | 80 | 162 | no | yes | 0.98 |
| InSDN | 2020 | virtualized | simulated | network | 56 | 0.2 | yes | no | 0.95 |
| IoT-IDS | 2019 | real | simulated | IoT | 8 | 0.3 | no | yes | 0.96 |
| ROSPaCe | 2023 | real | conventional | network, OS, ROS2 | 60 - 482 | 30.2 | yes | yes | ? |

*Table 1: Summary of the most famous dataset for intrusion detection, and comparison with ROSPaCe.*

Community Establishment (CSE) and the Canadian Institute for Cybersecurity. The dataset is implemented on Amazon Web Services and is prepared from a much larger network of virtualized client targets and attack machines[9]. The resulting dataset is composed of above 16 million data points gathered from 10 days of network traffic obtained with two synthetic profiles that simulate the user behavior and the attacker behavior.

*InSDN*[10] includes five attack types towards network services. To create normal traffic, the authors simulate HTTPS, HTTP, DNS, email, FTP, and SSH traffic. The dataset is generated using emulators, and traffic is collected using tcpdump on the host device.

*IoT-IDS*[35] is composed of real network traffic in an Internet of Things (IoT) application, with the inclusion of some generated attacks. The data collection involves two common smart home devices: an intelligent assistant and a Wi-Fi Camera, connected to the same wireless network.

In Table 1, we summarize the key aspects of the above datasets. The table reports i) the name of the dataset, ii) the release year, iii) the target environment, distinguishing from real systems or virtual networked systems, iv) if the normal (attack-free) operativity corresponds to the nominal activities of the deployed system (*conventional*) or is stimulated with synthetic user profiles (*simulated*), v) the type of features collected, identifying the architectural layers that are monitored, vi) the number of features monitored, vii) the number of data points, viii) the availability of attacks specific to the system, software, and application, in contrast to attacks to common network services, ix) if the dataset provides alternating patterns of normal behavior and periods under attack, to model the advent of an attacker in an attack-free system, or if the normal and attack behaviors are collected separately, and x) the accuracy of state-of-the-art supervised algorithms, obtained from past research works (accuracy of KDD Cup99, NSL-KDD, ISCX, CICIDS17, and IOT-IDS is from Thakkar et al.[43], accuracy of CICIDS18 is from Leevy et al.[9], accuracy of ADFA-LD is from Khandelwal et al.[14], accuracy of InSDN is from Negera et al.[17], and accuracy of ADFA-WD is from Khraisat et al.[11]).

Different from all the reviewed datasets, ROSPaCe provides features from multiple architectural layers (Linux, ROS2, network), and it is the first intrusion detection dataset that monitors ROS2 features. This focus on ROS2 stands also for the attack selection because we provide 3 different attacks specific to the exploitation of known ROS2 vulnerabilities. Furthermore, the dataset collects all traffic during the system's nominal execution, rather than simulating the normal usage of the system with a synthetic user profile.

Lastly, the ROSPaCe dataset describes the dynamic of an attacker trying to exploit the system, by providing observation in a time-series form where periods of normal execution and attack operation are alternated. By marking the start of attacker operations, we can measure the shortest time an attacker is discovered once it starts its attack sequence. This allows assessing IDS detection capabilities not only based on accuracy but also on a broader understanding of how fast the detector can recognize an attack. For example, notorious datasets like CICIDS17 are composed of long periods of normal operativity, and long periods where multiple attacks are injected. In this way, the scenario of an attacker entering the system and performing actions on an “attacker-free” system is insufficiently captured.

## The target system: SPaCe system and ROS2

The SPaCe system is designed to orchestrate and optimize public mobility, improving the user experience and preventing security violations. It performs surveillance of tram carriages using on-board cameras, to determine the operative conditions of the tram carriage in real-time. SPaCe analyses the video stream from cameras to establish the occupation of the vehicles and notify users and administrators. It also provides detection capabilities to request intervention in case of damage to the vehicle and its equipment.

The SPaCe system is structured into 3 basic components. A remote server orchestrates the fleet and dispatches information to the users and the administrators. The onboard architecture of each tram carriage collects, processes, and anonymizes the data from the onboard cameras so that it can be sent to the remote server for fleet optimization and coordination. Finally, user devices show the information to the end users, with different interfaces based on the user privileges.

In this paper, we focus on the tram carriage and its onboard architecture. We describe the architecture of the onboard system with the aid of Figure 1. Cameras installed inside the passenger carriage provide a video stream to the *In Camera Track* module. This module, in turn, makes the first processing of the video stream. The In Camera Track identifies the relevant elements within the scene and reconstructs their space-time consistency. This module is composed of a set of microservices: i) the *image manager* implements the data flow management: it interacts with the camera sensors by extrapolating the raw data and normalizing the data structure; ii) the *detector* allows the extraction of features relevant to the analysis of the status of vehicles and stations, e.g., the presence of people, object classification, and position assessment; iii) the *tracker* tracks a target object or person in successive frames.

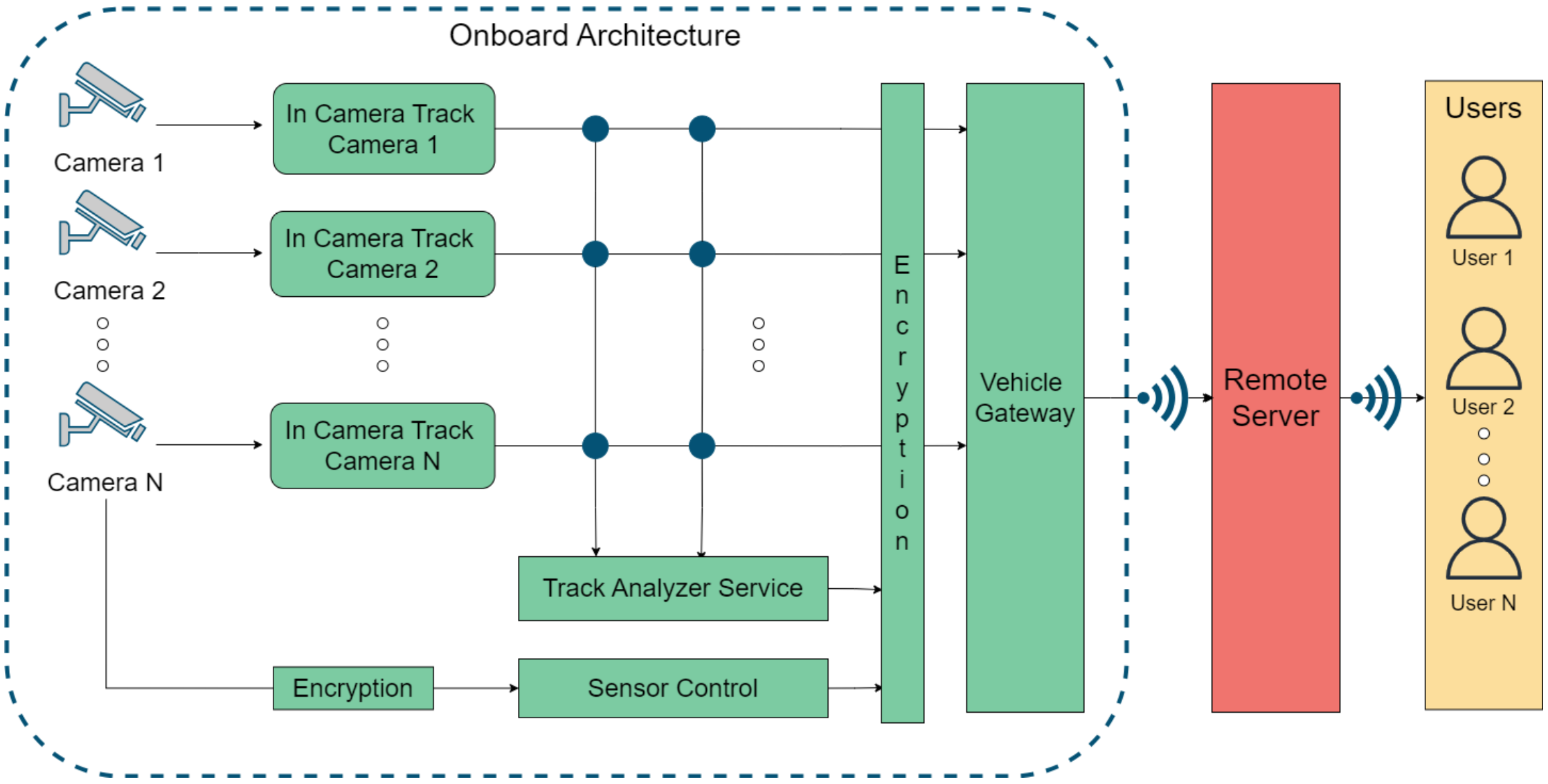

*Figure 1: The SPaCe architecture, with a focus on the on-board architecture. It is a cyber-physical embedded system where video cameras monitor the tram carriage, and the information is locally processed through machine learning algorithms. Elaborated and anonymized information is then transmitted to the remote server.*

The In Camera Track sends the elaborated information to the *Track Analyzer Service*. This module extrapolates the information useful to the SPaCe system, for example, if there are lost objects, broken items, dangerous situations, as well as status information on the capacity of the vehicle. More precisely, the Track Analyzer Service implements a set of microservices that provide insight into the traces detected by the trackers. Its main role is to reconstruct the information necessary for a posteriori analysis of the track. This module includes the following microservices: i) the *face and clothes attributes* extract the biometric information needed for facial recognition, and profiles users based on their clothes; ii) the *human keypoints* analyses the passenger movements a posteriori differentiating between legit or forbidden behavior; iii) the *object attributes* extract the attributes from the objects, to allow the subsequent recognition in a completely anonymous way. The information is then transmitted to the remote server. Lastly, the *Sensor Control* module encrypts and stores the raw data from the camera, which is available upon request.

The implementation of the on-board architecture relies on ROS2[1,2,3], an open-source software platform for developing robotic applications. It is largely applied in embedded cyber-physical systems[19,20], thanks to its ability to deploy and coordinate distributed microservices. ROS2 applications consist of independent computing processes called nodes that communicate in a publish/subscribe fashion, passing messages via a topic: a node publishes a message to a topic, and topic subscribers can utilize the message[2]. ROS2 also provides a communication paradigm alternative to publish/subscribe: a node can expose a service that can be exploited directly by another node, in a remote procedure call fashion. Unlike ROS, which relies on a centralized master process to handle communications, ROS2 exploits the standard Data Distribution Service[3] (DDS) which is completely distributed. The DDS introduces procedures to optimize the Quality-of-Service (QoS), especially to optimize the application for the available bandwidth and latency and to increase guarantees on message delivery. In SPaCe, the ROS2 *foxy* version is available as a middleware service on top of the Linux Ubuntu 20.04 operating system.

## Methods

We present the experimental setting and the steps to compose the dataset. The discussion is divided into three parts. First, we define our system and its setting, including the monitoring system, the data collection method, and the attacks towards the on-board architecture. Second, we present the data processing and labeling to create a unique dataset from traces collected by different monitors. Third, we describe the resulting datasets, and we reduce the set of features to provide a lighter version of the dataset.

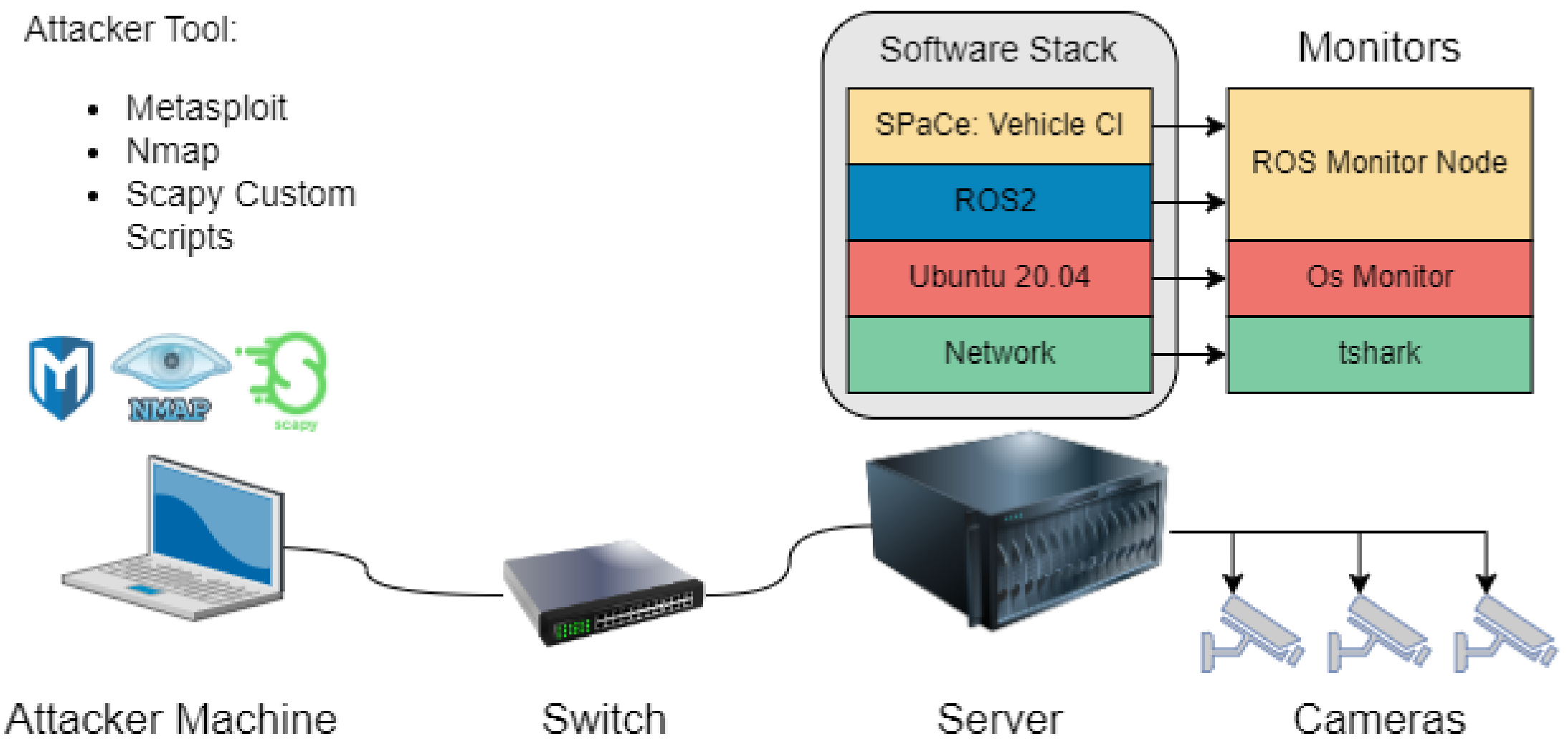

*Figure 2: The setting of our experimental campaign. The Ubuntu 20.04 Server runs ROS2, and it is connected to camera sensors. The attacker operates from an Ubuntu 20.04 laptop connected via a switch. The attacker tools are Metasploit, Nmap, and custom scripts implemented using the Scapy library. The software probes monitor data from three architectural layers: the network, the operating system, and the ROS2 middleware and microservices.*

### Part 1: Experimental Set Up

We describe the experimental setup with the aid of Figure 2. Our experimental set-up is composed of i) the onboard architecture, which is a server running Linux Ubuntu 20.04 and ROS2 foxy, and ii) the attacker machine, which is an Ubuntu 20.04 laptop connected to the onboard architecture through a switch. Monitors are installed on the server, to observe features from the network layer, the operating system layer, and ROS2. The data collection system requires the contribution of the attacker machine, and the server as will be discussed later.

**Attacks.** To run and implement the attacks, we rely on different tools installed on the attacker's machine, namely Nmap, Metasploit, and custom scripts we developed. The network mapper Nmap[8] is a free and open-source utility for network discovery and security auditing. Nmap uses raw IP packets to determine the available network hosts, the services (application name and version) and operating systems being executed on the target hosts, the type of packet filters and firewalls in use, and many other characteristics. We run Nmap version 7.93. Metasploit[4] relies on a large central repository of known exploits. The user can select an exploit from the central repository, and the payload to run after the breach. We rely on the PyMetasploit[6] library to integrate and run Metasploit in a Python program. Lastly, we select scripts that are specific to attack ROS2, and we customize them to fit our experimental setting. These scripts rely on Scapy[7], a powerful interactive packet manipulation library in Python. Scapy can forge or decode packets of a wide number of protocols, send them on the wire, capture them, and match requests and replies.

We select attacks using as reference an existing ROS2 threat model[36]. Consequently, we focus on attacks that exploit the external network, especially the network and the ROS2 DDS as they are highlighted among the main entry points. The attacks are from two categories, usually termed *discovery,* and *denial of service* (DoS). A discovery attack consists of collecting information about the target system, rather than exfiltrating sensible data or directly damaging the system. While they do not directly compromise the system, these attacks provide information that can be used by attackers to breach system defenses. Instead, DoS is aimed at reducing or interrupting the availability of a service exposed by the target system. An attacker can implement a DoS in multiple ways, targeting different parts of the system stack, still making the system's services unavailable.

Overall, we select 2 discovery attacks and 4 DoS attacks. Three attacks are generic and provided by state-of-the-art tools, and 3 are specific to the ROS2 DDS.

**Discovery attacks.** *Nmap Discovery* is mainly used to extrapolate the version of the operating system of the target machine. It relies on an internal database of fingerprints collected from a large variety of operating systems. It builds a fingerprint of the unknown system by analyzing the responses to crafted TCP and UDP requests. Then it searches for a complete or a partial matching of the fingerprint in the database. We use the Nmap OS Detection script[8] to run the attack with the command *nmap -O –osscan-guess*.

*ROS2 Reconnaissance* is a discovery attack specific to ROS2 that extrapolates information about the running nodes and their services. When the usage of a simple ROS2 query is not possible, it tries to exploit the vulnerabilities of the DDS by sending a specifically forged package. A vulnerable DDS will respond with a service message that lists the running nodes on the ROS2 architecture. We exercise the attack relying on the code from the Robot Hacking Manual[4].

**DoS attacks.** *Flooding Meta* is the Metasploit implementation of the common network-level SynFlood attack. It consists of sending many packets with the SYN bit =1 to request a new connection to the target machine. The target system replies with SYN-ACK packets, allocates resources to handle the connection, and waits for an ACK packet that confirms the connection. The attacker ends the interaction in this state causing the system to waste computational and memory resources. The script that implements the attack is *auxiliary/dos/tcp/synflood*[5] available in Metasploit.

*Flooding Nmap* is the Nmap implementation of a network-level DoS attack, that we exercise relying on the *ipv6-a-flood.nse*[8] script.

*ROS2 Node Crashing* exploits known vulnerabilities of some versions of the DDS which cause incorrect handling of ROS2 messages that do not respect the declared length. Like in buffer overflow attacks, the attacker can cause the execution of a piece of code hidden in the packet or can lead to a crash of the target ROS2 node. This attack is implemented with a custom Python script provided by the Robot Hacking Manual[4] and written using Scapy.

*ROS2 Reflection* exploits a vulnerability of the communication protocol used by the DDS. It consists of modifying the multicast address of a ROS2 node by sending a purposely crafted message. If successful, the attacker can redirect the node to listen to messages from a malicious server. The server produces continuous traffic to saturate the resources. The attack is implemented using Python and Scapy and the implementation is provided by the Robot Hacking Manual (RHM)[4].

**Monitors.** We describe the software probes that we set up to monitor the system execution. The probes are installed on the server as shown in Figure 2.

*Tshark*[33] is a well-known tool for monitoring network activity. It targets the network layer, and it is configured to collect all the packets transiting the network interface. The collected packets are logged in *.pcap* files for later processing. The list of the 452 monitored features is not described here for brevity, but it is available as a CSV file in the dataset repository.

The *OS Monitor* is a monitoring software we built to target the OS layer of a Linux machine. It collects all the information provided by the */proc* virtual filesystem[29], such as memory usage, kernel configuration, installed devices, the status of the CPU cores, and information specific to each process running on the OS. The OS monitor operates with a period of 200 milliseconds. In Table 2, we list the 25 features that we monitor from the operating system, along with their description.

| Indicator Name | Description |
|---|---|
| MemFree | The amount of RAM, in kilobytes, left unused by the system. |
| Buffers | The amount of RAM, in kilobytes, used for file buffers. |
| Cached | The amount of RAM, in kilobytes, used as cache memory. |
| Active | The total amount of buffer or page cache memory, in kilobytes, that is in active use. This is memory that has been recently used and is usually not reclaimed for other purposes. |
| Inactive | The total amount of buffer or page cache memory, in kilobytes, that are free and available. |
| SwapFree | The total amount of swap-free memory, in kilobytes. |
| Pgpgin | Number of pageins (since the last boot). |
| Pgpgout | Number of pageouts (since the last boot). |
| pgalloc_dma | Number of page allocations per zone (since the last boot). |
| Pgfree | Number of pagesfrees (since the last boot). |
| Pgactivate | Number of page activations (since the last boot). |
| Pgdeactivate | Number of page deactivations (since the last boot). |
| Pgfault | Number of minor page faults (since the last boot). |
| Pgmajfault | Number of major page faults (since the last boot). |
| Disk_Read | read operations executed on disk since the last observation. |
| Disk_Write | write operations executed on disk since the last observation. |
| Net_Received | Bytes of data received from the network since the last observation. |
| Net_Sent | Bytes of data sent through the network since last observation. |
| Tcp_Listen | Number of listening TCP sockets. |
| Tcp_Established | Number of active TCP sockets. |
| Tcp_Syn | Number of synchronized TCP sockets. |
| Tcp_TimeWait | Number of waiting TCP sockets. |
| Tcp_Close | Number of closed TCP sockets. |
| nr_active_file | Number of active files |

*Table 2: Monitored features of the virtual filesystem /proc/.*

The *ROS Monitor* is a custom node that we developed to monitor the ROS2 layer. The node subscribes to each topic published by the active nodes. The ROS Monitor scans periodically for new nodes to monitor changes in the system that may occur over time. The lists of monitored nodes include the */rosout* node which is the default logger of the ROS2 architecture and collects logging data from the active nodes. Implementing the ROS Monitor node avoids directly using the */rosout* logging operation, which would create a *bag file* to save every ROS2 message. Producing a bag file is invasive and burdensome, meaning that it is too intrusive for monitoring purposes. The ROS monitor operates with a period of 200 milliseconds. In Table 3 we report the 5 features obtained from the monitor along with their descriptions.

| ROS2 Feature Name | Description |
|---|---|
| src_topic | Indicates the source node of a specific message. |
| subscribers_count | Number of subscribers to a topic. |
| publisher_count | Number of publishers to a topic. |
| msg_type | Type of a ROS2 message. |
| msg_data | The header of a ROS2 message. |

*Table 3: Monitored features of the ROS Monitor Node. It collects information from each topic including /rosout which is the default logging system in ROS2.*

**Experimental campaign.** Once the onboard server and the monitors are up and running, we execute a custom Python script on the attacker machine, to coordinate and automate the experimental campaign. The script, schematized in Figure 3, iteratively performs the following actions, for the number of attacks:

1. Select an attack *x* that was never selected before.
2. Wait 30 seconds, without performing any action. In this period, the target system is behaving normally, in an attack-free scenario.
3. Inject the attack *x*, for a maximum temporal duration *t*.
4. Stop injecting *x* and keep the system in a hiatus state for 10 seconds. This period is set to allow the system to restore a clean state. We expect that after this period, the effect of the attack will no longer manifest on the monitored features. In case services are crashed or unable to recover, the system is rebooted, and the experimental campaign restarts from step 2.
5. Repeat steps 2, 3, and 4 for 400 iterations. This means that 400 sequences of normal operations and successive injections of attack *x* will be logged.
6. Wait 90 seconds to restore a clean system state, then restart from 1.

We set the temporal duration *t*=60 seconds for Flooding Meta and Flooding Nmap attacks, after which the attack is interrupted. The duration of the other attacks depends on many factors and can change at each run, with a timeout of *t*=60 seconds. The waiting time of steps 4 and 6 is introduced to allow system recovery after an attack, such that the value of the monitored features is no longer altered by the attack.

All the steps above are commanded by the attacker machine, and locally timestamped and logged.

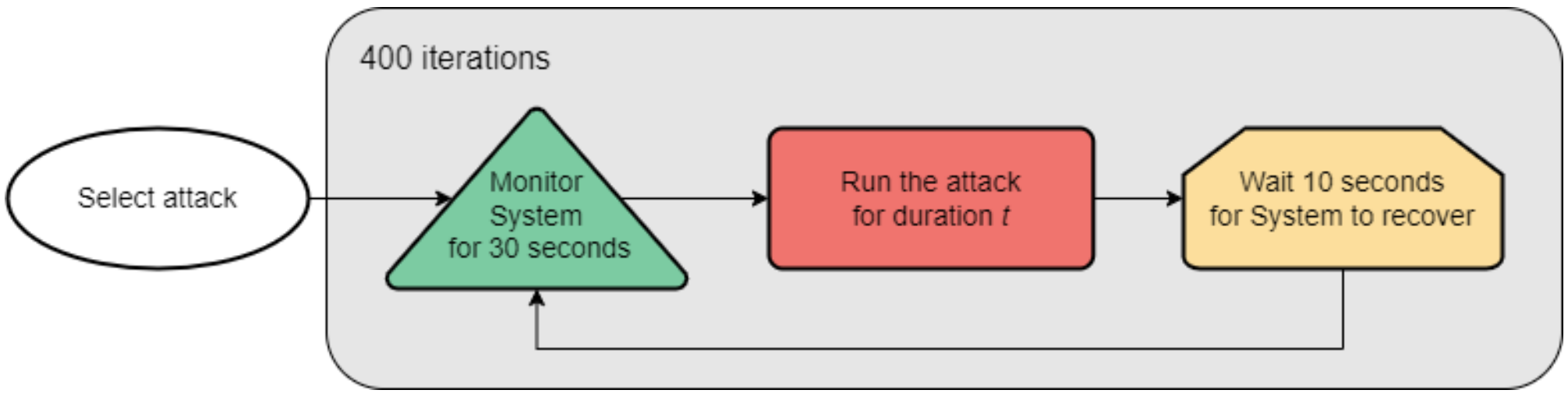

*Figure 3: The script that automates the execution of the experimental campaign.*

## Part 2: Data Processing and Labeling

At the end of the experimental campaign, we obtain three logs from our three monitors for a total of 27.3 GB. The ROS monitor and the OS monitor log in CSV format, while Tshark records *pcap* files. We export the pcap files collected from Tshark in a *json* format, generating multiple files of about 10 GBs each, and then we process the json files with the Pandas[21] library to obtain CSV files.

To merge the CSV files from the 3 different monitors, we use the timestamp as a key, and as a reference the CSV file obtained from Tshark. The reason is that Tshark, differently from the other monitors which logs with a period of 200 milliseconds, is not configurable with a sampling interval and collects any packet. Consequently, the Tshark CSV file contains observations at much shorter intervals compared to the other monitors. The data merging process is illustrated in Figure 4. We concatenate each row from the Tshark CSV file with rows of the OS monitor and ROS monitor, if their timestamps differ by less than 100 milliseconds (ms). Conversely, if the difference between the timestamps is greater than 100 ms, the rows are not concatenated. This algorithm works given the logging period of 200 ms for both the ROS monitor and the OS monitor. Still, due to unavoidable jitters in the logging period, in some cases, a row may report values for Tshark features, while the features from one or both the other two monitors are left blank. Due to computational constraints and the file size, we perform this operation by splitting the CSV file into batches.

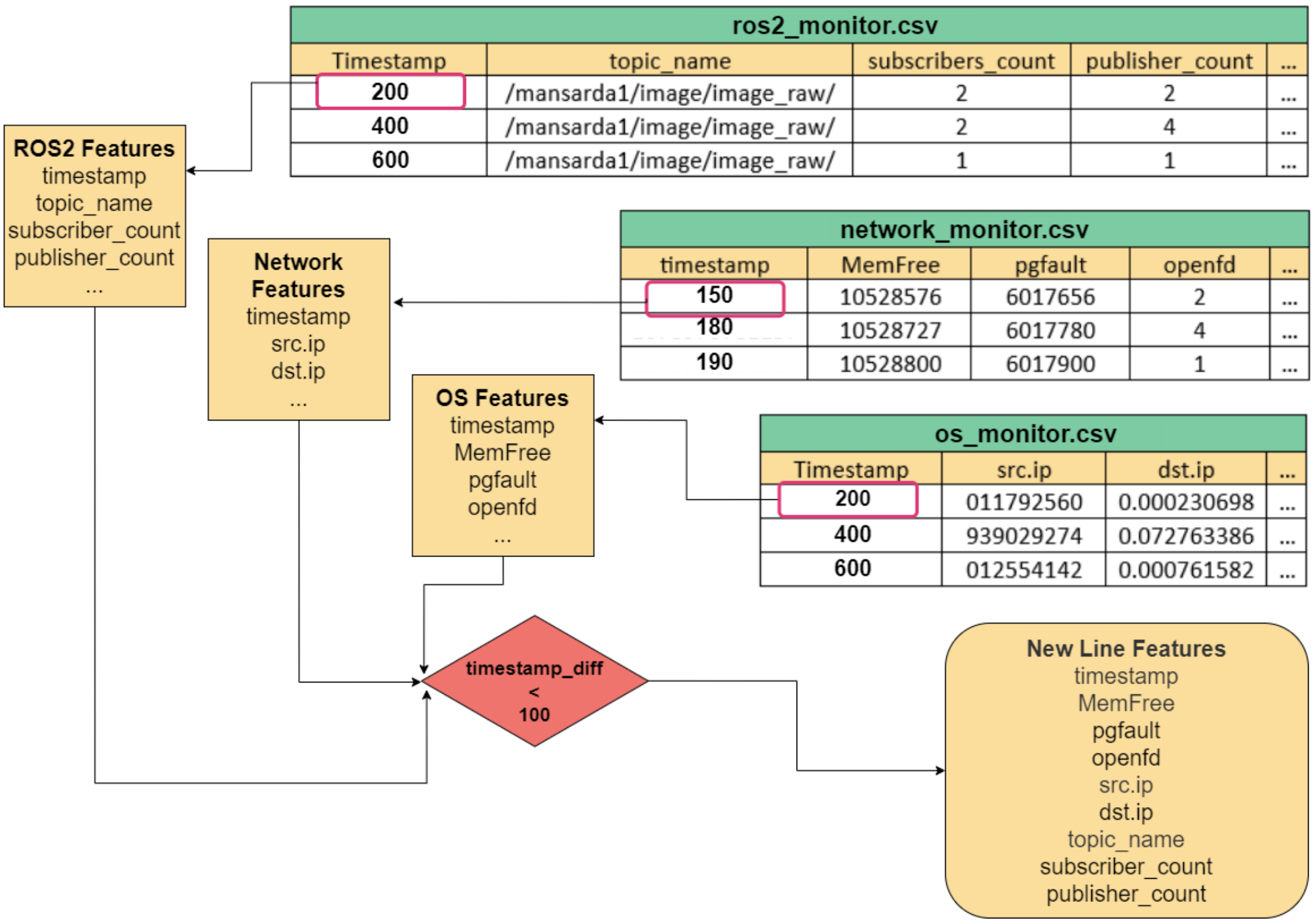

*Figure 4: Data merging procedure to compose a unique CSV file starting from the logs produced by the three monitors.*

To label each data point (each row) of the resulting CSV file, we rely on the log of the attacker machine. The attacker machine locally logs a timestamp at each step of the experimental campaign. We use these timestamps to identify i) the start of the 30 second period of normal operation; ii) the start of the successive attack period; and iii) the termination of the attack period. This way, we can identify data points produced in the absence of attacks, that we label as *normal*, and data points produced when a specific attack is exercised, that we label with the name of the attack. Data collected during the restoration periods are discarded.

Figure 5 schematizes the labeling procedure. We label the observations between the “observe” and the “start” timestamps as normal, and we label the observations between the "start" and the "end" timestamps with the name of the attack. We discard the observations between the “observe” and “start” labels because they are collected after an attack, while we are allowing the system to recover. With this procedure, we identify observations collected during the temporal execution of the attacks, with the only approximation of time synchronization between the two nodes, which are synchronized to a global time server using the Network Time Protocol[34].

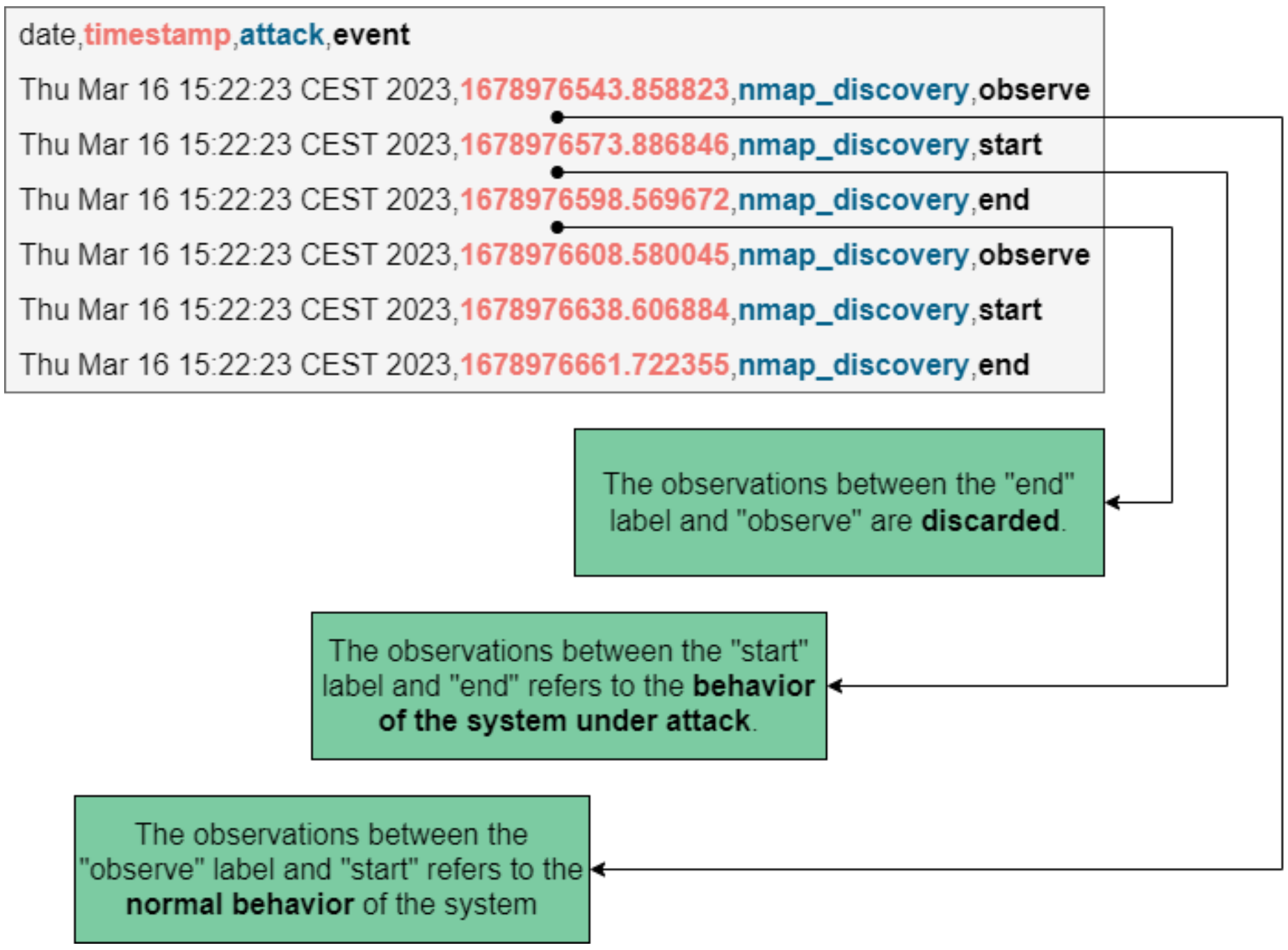

*Figure 5: The observations between "start" and "end" are labeled with the attack name, while the observations between "observe" and "start" are labeled as normal. We discard the observations between the "observe" and "start" labels because they are collected while the system is recovering to a clean state.*

We remark that we do not indicate when the effect of a given attack manifests on the monitored features, but we indicate when the attack session starts. In other words, while we can identify when the first attack packet is launched, we cannot tell when (or if) the attack will affect the value of the monitored features and consequently may be visible to anomaly detection.

Last, we observe that the data logged by Tshark is inconsistent, meaning that the number of features logged varies from one data point to another. To homogenize the dataset and present a uniform set of features, we concatenate the CSV file batches by applying an inner join operation on the columns. This way, we select only the features that are in common between all files. Further, we delete: i) features whose values are linear indexes, that simply identify the order of data points; ii) duplicate features; iii) features with only 1 value. At the end of this process, the number of features is 482, excluding the label.

## Data Records

**ROSPaCe dataset**. The final shape of the ROSPaCe dataset is 30 247 050 data points and 482 columns excluding the label. The features are 25 from the OS monitor, 5 from the ROS Monitor Node, and 422 from Tshark. The dataset is encoded in the *ROSpace_complete.csv* file for a total of 40.5 GB. The dataset contains about 23 million attack data points and above 6.5 million normal data points (78% attacks, 22% normal). We specify in Figure 6 the number of observations collected. Reasonably, DoS attacks have the highest number of attack data points. Instead, the low number of data points of *ROS2 Reflection* and *ROS2 Node Crashing* is because the attacks lead to the failure of one or more ROS2 nodes, in some cases including the monitors. Note that these cases required a complete restart of the system and an interruption of the experimental campaign. This is also confirmed in Table 4, where the left part describes the temporal duration of the attack sequence, for each attack type, in terms of average, minimum, and maximum duration, while the right part describes the average, minimum, and maximum number of data points in the attack sequence.

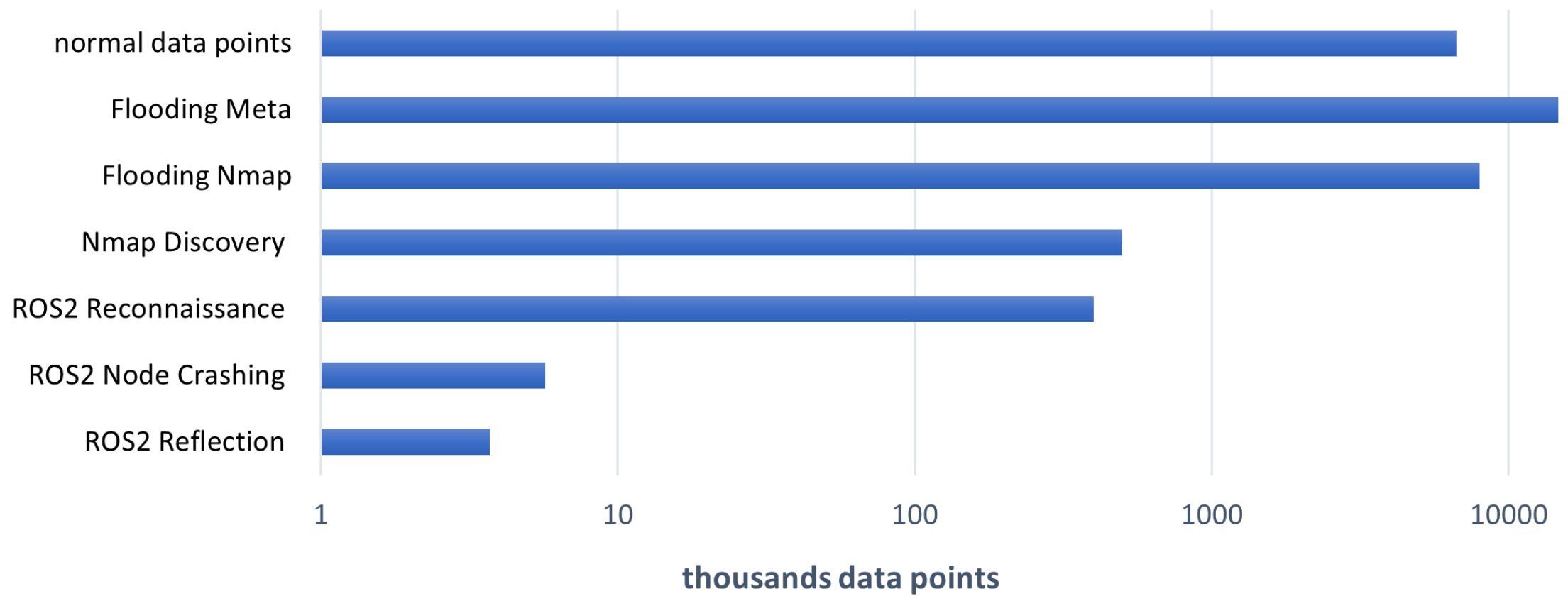

*Figure 6: Number of attack data points and normal data points. The x-axis is in logarithmic scale.*

| attack | average duration | minimum duration | maximum duration | average length | minimum length | maximum length |
|---|---|---|---|---|---|---|
| flooding meta | 60.97 | 59 | 160 | 43 303 | 812 | 171 799 |
| nmap discovery | 22 | 20.45 | 24.66 | 1 377 | 1 180 | 8 248 |
| flooding nmap | 29.86 | 29.80 | 29.95 | 398 655 | 467 044 | 40 534 |
| ROS2 node crashing | 1.05 | 0 | 40.23 | 51 | 2 | 150 |
| ROS2 reconnaissance | 13.14 | 9.36 | 60.16 | 1 907 | 1 295 | 4 873 |
| ROS2 reflection | 3.69 | 0 | 40.20 | 24 | 1 | 99 |

*Table 4: Left: average, minimum, and maximum temporal duration of attack sequences, in seconds. Right: average, minimum, and maximum number of data points in an attack sequence.*

**ROSPaCe reduced dataset.** We provide a lightweight version of the ROSpace dataset by selecting the best-performing 60 features. This includes the 30 features from the OS monitor and the ROS monitor and the 30 best-performing features from Tshark.

We proceed by inspecting the dataset to identify and remove those that may reduce the model's ability to generalize, potentially leading to a degradation in detection performance. For example, these are features that specify the source or destination at different layers of the protocol stack. Then, we reduced the Tshark features by selecting the 30 most important ones using an Extremely Randomized Trees[30] meta-estimator. This meta-estimator fits a fixed number of randomized decision trees on a subset of samples of a dataset and uses averaging to select the features that contribute the most to the classification, i.e., identifies the features that are mostly used to make classification decisions.

The ROSPaCe reduced dataset is encoded in the *ROSpace_reduced.csv* file for a total of 13.5 GB.

## Code Availability

The software code to run the entire experimental campaign can be downloaded from the *codes* folder of our open software repository[31]. Each subfolder corresponds to a specific activity of the dataset creation procedure reported in this paper.

The ROSPaCe and ROSPaCe reduced datasets are available in a public repository and can be freely downloaded. The repository contains a folder for each of the datasets, and an additional CSV file that enlists the features included.

## Usage Notes

The steps delineated in this section are aimed at arranging the data into a structure that is well-suited for efficient and effective utilization. The steps are implemented in two *Jupyter* notebooks available in our software repository[31], and they can be executed straightforwardly only at the cost of setting the correct path to the dataset file.

First, we reduce to numeric all the features with mixed data types (e.g., alphanumeric, and numeric values). Second, we replace *nan* values and infinite values with -1 as they are usually not accepted by ML algorithms. Then, we map attack labels to numeric values: to distinguish between normal data points and attack data points, we just need binary classification and consequently, we transform each attack label to 1 and each normal label to 0. Third, we convert string values to numbers using categorical encoding. This technique assigns a unique number to each unique string value of a feature. After implementing this preprocessing, the dataset can be used for two purposes: to measure the ability to distinguish between normal and attack data points and to measure the attack latency. Each approach relies on the application of additional preprocessing steps, described respectively in the Usage Notes-A and Usage Notes-B below.

### Usage Notes – A

Our focus is to distinguish between normal and attack data points: therefore, we drop all the time- and sequence-related features, and we shuffle the dataset. Timestamps or time-related data in the dataset can inadvertently serve as labels, potentially leading to misleading results. Very practically, the steps are:

1. Drop all columns containing timestamps or that may contain information on data sequences.
2. Shuffle the dataset.
3. Divide the dataset into train and test sets by applying a 60/40 split.
4. Train and test a binary classifier, compute the appropriate metrics (see the Technical Validation below), and visualize results.

### Usage Notes – B

In this scenario, we shift our focus to investigate how long an attacker is undetected. The notion of time, in this case, is central: we want to train the ML detector to make decisions based on system evolution through time rather than deciding on individual data points. Therefore, the time-related features must be included in the dataset to relate subsequent data points. As the dataset is already organized in alternating blocks of normal behavior and attack scenarios, the only steps needed are the following:

1. Identify all the pairs (normal sequence, attack sequence) in the dataset. We call each of these pairs a *block*.
2. 60% of randomly selected blocks are assigned to the train set, and the remaining 40% to the test set. This way we train the ML detector with blocks of ordered data points, each block containing approximately 30 seconds of normal data, that are followed by a variable amount of attack data.
3. Train and test a binary classifier, compute the appropriate metrics (see the Technical Validation below), and visualize results.

## Technical Validation

In this section, we show that the dataset is well-formed and can be used for intrusion detection purposes. We use the reduced dataset (*reduced_dataset.csv*) as it is easier to manage.

### Validation with Usage Notes - A

As intrusion detectors, we train and run two different popular machine learning algorithms, namely, the unsupervised Isolation Forests[25], and the supervised XGBoost[27].

We evaluate the performance using the accuracy, the ROC curve, and the precision-recall curve. Given True Positives (TP) and True Negatives (TN) as the correctly predicted values, and False Positives (FP) and False Negatives (FN) as the misclassified events, the accuracy measures the correct predictions in a classification task, and it is defined as A = (TP

+ TN) / (TP+TN+FP+FN). The ROC curve plots the probability that the model will correctly rank a randomly chosen positive instance higher than a random negative one[23,24]: it represents the trade-off between recall and False Positive Rate, where recall is R= TP / (TP+FN), and the false positive rate is FPR = FP / (TP + FP). Further, we use the Precision-Recall curve which calculates the trade-off between precision and recall for different detection thresholds, where precision is P = TP / (TP+FP).

The Isolation Forest algorithm achieves accuracy A=0.875 with FPR=0.057. The recall score is R=0.895, with a related precision P=0.942. XGBoost shows drastically better results showing accuracy A=0.996 with FPR=0.0018. The recall score is R=0.997 with a corresponding precision of P=0.998. In Figure 7, we detail the ROC curve and the precision-recall curve for Isolation Forest and XGBoost. Their capability to detect attacks validates the usability of our dataset: attacks have a visible effect on the monitored system features, and the dataset can be used to train an anomaly-based intrusion detector. In addition, we observe that the ROC and precision-recall curves show opposite and regular trends, assuring that the accuracy result is not biased by inter-class imbalance.

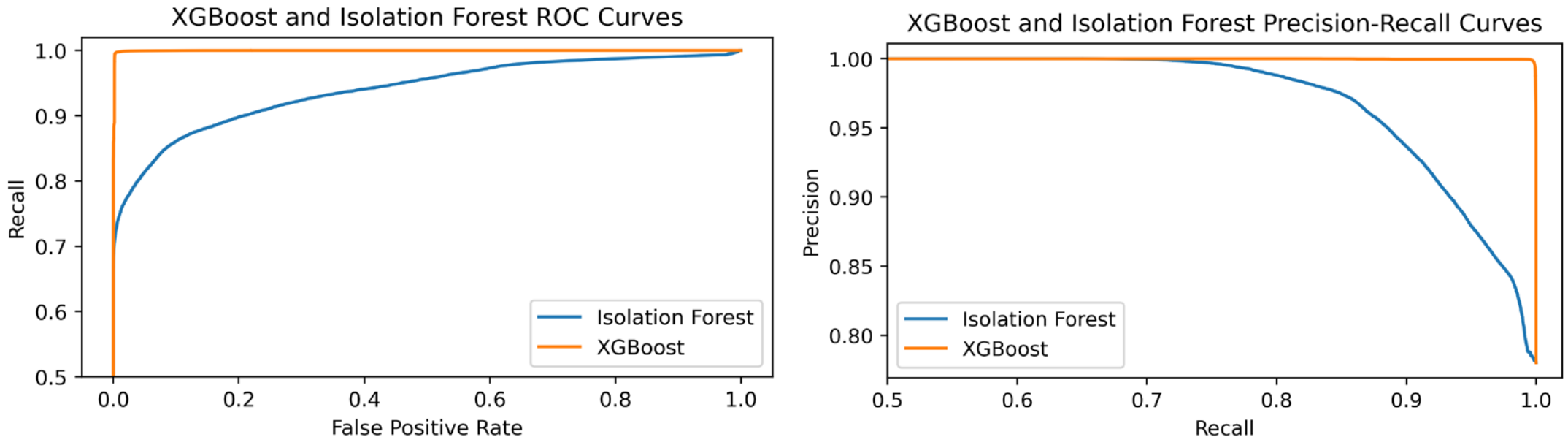

*Figure 7: ROC curve and Precision-Recall curve of XGBoost and Isolation Forest on the ROSPaCe reduced dataset.*

## Validation with Usage Note – B

We take advantage of the ROSPaCe characteristics to evaluate the attack latency, i.e., measure how long an attack will be perpetrated in the system before being detected. The metrics from Usage Note - A may present a misleading understanding of the actual capability to defend from attacks. For example, accuracy seems satisfactory, but the detector may have missed the initial part of the attack, which still might be sufficient to reach determined attack goals.

We resort to the organization in blocks of the test set produced in Usage Note-B. For each block, we keep track of two data points: i) the first attack data point, and ii) the first attack data point that is correctly classified. The attack latency is the distance between the two data points: it can be measured as a time interval, or as the number of data points that occurred between the two attack data points.

In addition, we also are interested in understanding the number of false positives, which is identified by the FPR. A high FPR means many false alarms, which ultimately make the system unusable: consequently, we decide on a maximum acceptable FPR, and we compute the recall and the attack latency. In this paper, we set the maximum FPR=$10^{-4}$, which means that during the normal operativity, one data point every $10^4$ data points is erroneously classified as an attack. We are aware that, in practice, the FPR threshold should be agreed upon with SPaCe engineers and security managers, but the deployment of an IDS is outside the scope of the paper.

As an algorithm, we use XGBoost to classify each data point in a block. Time series algorithms, like long short-term memory (LSTM[26]) recurrent neural networks, could also be exploited; however, a comparison of the most suitable algorithm for ROSPaCe is outside the scope of the paper.

We report the performances of XGBoost in Table 5, Table 6, and Table 7. Table 5 describes the attack latency in terms of seconds elapsed from the start of the attack until the first detection. Table 6 presents the number of consecutive false negatives (misclassified attack data points) from the start of the attack sequence until the first detected attack. Table 7 reports the number of blocks where there is at least one detected attack.

Aggregating the results from the different tables, is evident that we can detect Flooding meta, Nmap discovery, and Flooding nmap, while we show a very low capability in detecting the three ROS2 attacks. These attacks are less represented in the dataset having shortest sequences with a smaller number of packages, because they lead the system in a failed state.

We observe that all the Flooding nmap blocks are detected (Table 7), with a very short detection latency: the average latency is 0.18 seconds (Table 5), during which 906 packets are received (Table 6). Somehow good results are obtained with nmap discovery, where we can detect at least one attack data point in 99% of the attack sequences. However, attack latency is significantly higher (6.74 seconds), but it is a direct consequence of the differences between the two attacks. The other attacks are more difficult to detect, and especially the ROS2 specific attacks, which have a very low detection ratio in Table 7. This suggests that the effect of ROS2 attacks on the monitored features is rather different from one block to another, and XGBoost is not able to generalize sufficiently to capture this difference. While this may look like a concern for the development of efficient intrusion detectors, these insights can serve as a foundation for developing a new approach for training and fine-tuning an intrusion detection system.

| attack | average detection latency | standard deviation | minimum detection latency | maximum detection latency |
|---|---|---|---|---|
| flooding meta | 13.08 | 19.58 | 0.18 | 126.80 |
| nmap discovery | 6.74 | 4.14 | 0 | 13.16 |
| flooding nmap | 0.18 | 0.09 | 0.04 | 0.35 |
| ROS2 node crashing | 0.05 | 0.04 | 0.01 | 0.09 |
| ROS2 reconnaissance | 5.63 | 2.58 | 2.33 | 8.46 |
| ROS2 reflection | 20.10 | 28.38 | 0.03 | 40.17 |

*Table 5: Attack latency, in seconds. We provide the minimum value, the maximum value, and the standard deviation.*

| attack | average detection index | standard deviation | minimum detection index | maximum detection index |
|---|---|---|---|---|
| flooding meta | 2 111 | 3 196 | 14 | 16 245 |
| nmap discovery | 167 | 350 | 0 | 3 613 |
| flooding nmap | 906 | 3 | 4 229 | 1 637 |
| ROS2 node crashing | 29 | 11 | 16 | 37 |
| ROS2 reconnaissance | 943 | 424 | 397 | 1 420 |
| ROS2 reflection | 22 | 6 | 18 | 27 |

*Table 6: Attack latency, in terms of the number of data items that occurred before detection. We provide the minimum value, the maximum value, and the standard deviation.*

| attack | total blocks | blocks with detections | detection ratio |
|---|---|---|---|
| flooding meta | 105 | 65 | 0.62 |
| nmap discovery | 162 | 160 | 0.99 |
| flooding nmap | 8 | 8 | 1.0 |
| ROS2 node crashing | 40 | 3 | 0.07 |
| ROS2 reconnaissance | 78 | 5 | 0.12 |
| ROS2 reflection | 78 | 2 | 0.02 |

*Table 7: Ability to detect at least one attack in a block. We report the total number of blocks in the test set, where at least an attack is detected, and the corresponding detection ratio.*

## Acknowledgments

This work has been partially supported by the Regione Toscana with the project POR CREO SPaCe, by project SERICS (PE00000014) under the MUR National Recovery and Resilience Plan funded by the European Union - NextGenerationEU, by the PRIN 2022 project FLEGREA (B53D23012930006) funded by the Italian Ministry of University and Research.

## Author contributions

Andrea Ceccarelli: experiments design; data collection; data processing; data analysis; writing. Cosimo Cinquilli: experiments design; implementation; data collection; data processing. Simone Nardi: experiments design; data collection; data processing; data analysis. Tommaso Puccetti: experiments design; implementation; data collection; data processing; data analysis; writing. Tommaso Zoppi: experiments design; data processing; data analysis.